\documentclass[epjST]{svjour}

\usepackage{graphics}

\begin{document}

\title{Interlayer tunneling spectroscopy of graphite at high magnetic field oriented parallel to the layers}

\author{Yu. I. Latyshev\inst{1} \and A. P. Orlov\inst{1} \and P. Monceau\inst{2} \and D. Vignolles\inst{3} \and S. S. Pershoguba \inst{4} \and Victor M. Yakovenko\inst{4}}

\institute{Kotelnikov Institute of Radio Engineering and Electronics of RAS, Mokhovaya 11-7, Moscow, 125009 Russia \and Institut N\'eel, CNRS, 25 rue des Martyrs BP 166, Grenoble, 38042, France \and Laboratoire National des Champs Magn\'etiques Intenses - Toulouse, CNRS, 143 avenue de Rangueil, Toulouse, 31400, France \and Condensed Matter Theory Center, Department of Physics, University of Maryland, College Park, Maryland 20742-4111, USA}

\abstract{
Interlayer tunneling in graphite mesa-type structures is studied at a strong in-plane magnetic field $H$ up to 55~T and low temperature $T=1.4$~K. The tunneling spectrum $dI/dV$ vs.\ $V$ has a pronounced peak at a finite voltage $V_0$. The peak position $V_0$ increases linearly with $H$.  To explain the experiment, we develop a theoretical model of graphite in the crossed electric $E$ and magnetic $H$ fields. When the fields satisfy the resonant condition $E=vH$, where $v$ is the velocity of the two-dimensional Dirac electrons in graphene, the wave functions delocalize and give rise to the peak in the tunneling spectrum observed in the experiment. }
\maketitle

\section{Introduction} \label{sec:0}

Over the last decade, interlayer tunneling spectroscopy was developed to measure energy gaps in high-temperature superconductors and charge-density-wave materials \cite{ref1,ref2}. It was also used to study magnetic-field-induced charge-density waves in NbSe$_3$ \cite{ref3} and graphite \cite{ref4}. This latter effect has orbital origin and generally exists for a magnetic field oriented perpendicularly to the conducting layers. In contrast, the interlayer tunneling in a parallel magnetic field can be utilized to obtain information about the in-plane energy spectrum of the carriers in the layers \cite{ref5}. Tunneling experiments between two graphene sheets in the graphene/insulator/graphene heterostructure were reported in \cite{ref6}. The effect of the in-plane magnetic field was studied theoretically for graphene multilayers \cite{ref7} and for a thin film of a topological insulator theoretically \cite{ref7b,ref8} and experimentally \cite{ref8b}. Here we present an experimental and theoretical study of graphite in strong in-plane magnetic and out-of-plane electric fields.

\section{Experimental Results} \label{sec:1}

\begin{figure}
\centering
\resizebox{0.4\columnwidth}{!}{\includegraphics{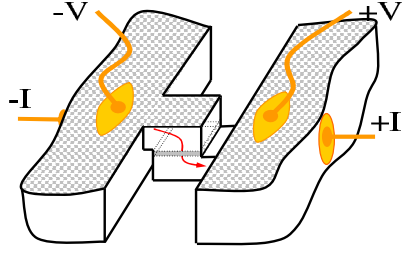}}
\caption{Schematic view of the mesa-type structure.}
\label{fig:1}     
\end{figure}

\begin{figure}
\centering	
\resizebox{0.4\columnwidth}{!}{\includegraphics{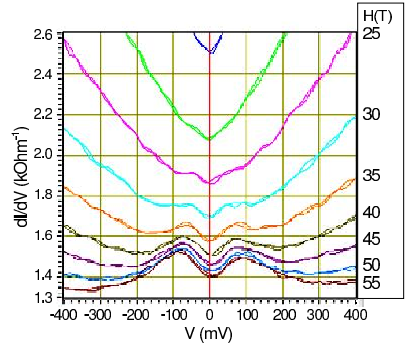}}
\caption{Interlayer tunneling spectra of the graphite mesa-structure at various in-plane magnetic fields at $T=1.4$K.}
\label{fig:2}     
\end{figure}

\begin{figure}
\centering
\resizebox{0.4\columnwidth}{!}{\includegraphics{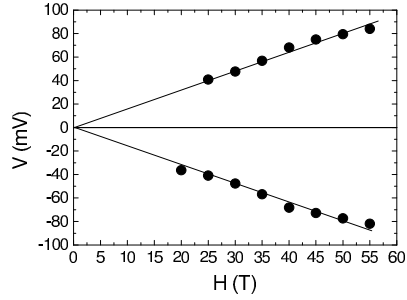}}
\caption{Dependence of the voltage $V_0$ of the tunneling conductance peak on the in-plane magnetic field $H$.}
\label{fig:3}     
\end{figure}

Mesa-type structures were fabricated by double-sided etching of thin graphite flakes using focused ion beam \cite{ref9}. We used high-quality natural graphite of NGS Naturgraphit GmbH Co. The mesa is shown schematically in Fig.~\ref{fig:1}. Electric current flows from one part of the crystal into another across the layers in the region of the mesa. The mesa is typically of 1 micron size and contains a few tens of graphene layers. Because of a very high interlayer conductivity anisotropy ($\sigma_\perp/\sigma_\parallel \approx 10^4$ at low temperatures), the applied voltage $V$ drops mostly on the mesa. The experiment was performed in pulsed magnetic fields at the National Laboratory for High Magnetic Fields in Toulouse. We developed a system of fast data acquisition, which allowed us to collect about $10^3$ $I$-$V$ characteristics, each containing $10^3$ points within each magnetic field pulse of $400$~ms duration \cite{ref3}. Figure~\ref{fig:2} shows a set of $dI/dV$ spectra vs. $V$ for various in-plane magnetic fields. For magnetic fields greater than $20$~T, local maxima start to develop for both polarities of the bias voltage symmetrically. Both the height and the voltage $V_0$ of the peaks grow with the magnetic field $H$. Figure~\ref{fig:3} shows that the dependence of $V_0$ vs.\ $H$ is close to linear.

\section{Theoretical Discussion} \label{sec:2}

\begin{figure}
\centering
\resizebox{0.4\columnwidth}{!}{\includegraphics{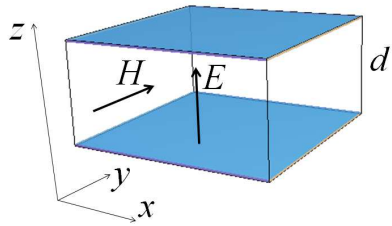}}
\caption{Schematic view of two graphene layers and the directions of the applied electric and magnetic fields.}
\label{fig:4}     
\end{figure}

In this section, we consider a model of graphite in crossed electric and magnetic fields in the geometry shown in Fig.~\ref{fig:4}. The magnetic field $\vec H=H\hat {\vec y}$ is parallel to the layers, while the electric field $\vec E=E\hat{\vec z}$ is perpendicular, and the distance between the layers is $d$. We shall demonstrate that, when the resonance condition for the fields
\begin{equation}  \label{eq1}
	E=vH
\end{equation}
is achieved, the electronic wave functions become delocalized in the $z$ direction and give rise to a peak in the interlayer differential conductance, as observed experimentally in Figs.~\ref{fig:2} and \ref{fig:3}.  The parameter $v=10^6$ m/s in Eq.~(\ref{eq1}) is the velocity of the two-dimensional Dirac electrons in graphene. 

\begin{figure}
\centering	
\resizebox{0.9\columnwidth}{!}{\includegraphics{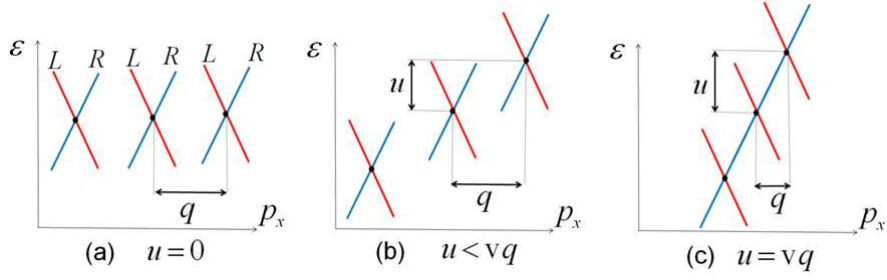}}
\caption{Schematic energy spectra of graphite in crossed magnetic and electric fields (represented by $q$ and $u$) in the limit $t\rightarrow 0$ at different values of $u$. (a) The Dirac cones have the same energy for zero electric field at $u=0$. (b) With an increase of $u$, the Dirac cones shift vertically. (c) When the resonant condition $u=vq$ is met, the R-branches of the Dirac cones align and become degenerate.}
\label{fig:5}     
\end{figure}

The Lorentz force from the in-plane magnetic field $H$, acting during interlayer electron tunneling, shifts the in-plane momentum $p_x$ of the electrons on the $n$-th graphene layer to $p_x-qn$, where $q=eHd$ \cite{ref7,ref8}. The uniform out-of-plane electric field $E$ changes the potential energy on the $n$-th layer by $un$, where $u=eEd$. Thus, using the approach of Ref.~\cite{ref7}, the Schr\"odinger equation for graphite in the crossed electric and magnetic fields can be written as 
\begin{equation}  \label{eq2}
	v\sigma_x(p_x-qn)\Psi_n+tI^A(\Psi_{n-1}+\Psi_{n+1}) = (\varepsilon-un)\Psi_n.
\end{equation}
Here, the Pauli matrix $\sigma_x$ acts on the spinor wave function $\Psi_n=(\psi_n^A,\psi_n^B)$, which has components on the A and B sublattices of the $n$-th graphene layer. The matrix $I^A=(1+\sigma_z)/2$ and the amplitude $t=0.4$~eV describe the interlayer coupling between the carbon atoms, which lie on top of each other in the Bernal-stacked graphite lattice. For simplicity, we set the in-plane momentum component $p_y=0$ below; however all calculations can be easily generalized for $p_y\neq 0$. In the limit of zero interlayer coupling $t\rightarrow 0$, the energy spectrum is given by a series of Dirac cones
\begin{equation} 	\label{eq3}
    \varepsilon_m = \pm v(p_x-qm)+um, 
\end{equation}
which correspond to the wave functions localized on $m$-th graphene layer.  Evolution of the spectrum (\ref{eq3}) with the increase of the electric field is qualitatively illustrated in Fig.~\ref{fig:5}. Each Dirac cone consists of two branches corresponding to the left-moving and right-moving electrons, referred to as the L-movers and R-movers in the rest of the paper and labeled in Fig.~\ref{fig:5}(a). For $u=0$, the Dirac cones are shifted horizontally by $q$, as shown in Fig.~\ref{fig:5}(a). For $u\neq 0$, the Dirac cones are also shifted vertically, as shown in Fig.~\ref{fig:5}(b). When the fields $E$ and $H$ satisfy the resonant condition $u=qv$, which is equivalent to Eq.~(\ref{eq1}), the right-moving branches of the spectra align and become degenerate, as shown in Fig.~\ref{fig:5}(c).

For a more detailed analysis, let us perform a unitary transformation $\Psi_n = e^{-i\sigma_y\pi/4}\Psi_n'$ to the basis of the L and R movers, $\Psi_n' =(\psi_n^R,\psi_n^L)$. Then, Eq.~(\ref{eq2}) becomes
\begin{equation}  \label{eq4}
	v\sigma_z (p_x-qn)\Psi'_n+\frac{t}{2}(1-\sigma_x) (\Psi'_{n-1}+\Psi'_{n+1}) = \left(\varepsilon-un \right)\Psi'_n.
\end{equation}
If we temporary neglect the matrix $\sigma_x$, then Eq.~(\ref{eq4}) decouples for the R and L modes:
\begin{eqnarray}
	&& \left[ vp_x+n(u-qv) \right] \psi_n^R+\frac{t}{2}(\psi_{n-1}^R+\psi_{n+1}^R)=\varepsilon \psi_n^R, \label{eq5}   \\
	&& \left[ vp_x+n(u+qv) \right] \psi_n^L+\frac{t}{2}(\psi_{n-1}^L+\psi_{n+1}^L)=\varepsilon \psi_n^L. \label{eq6}
\end{eqnarray}	
Solutions of Eqs. (\ref{eq5}) and (\ref{eq6}) can be enumerated by an integer $m$ \cite{ref7}, so that the eigenvalues are given by Eq. (\ref{eq3}), and the eigenfunctions are
\begin{equation}  \label{eq7}
	\Psi_n^{R,L}=J_{n-m} \left( \frac{-t}{u \mp qv} \right),
\end{equation}
where  $J_m (x)$ is the Bessel function. The spectrum (\ref{eq3}) is given by the two sets of curves $\varepsilon_m^R=vp_x+m(qv-u)$ and $\varepsilon_m^L=-vp_x+m(qv+u)$, which have linear dispersion in $p_x$. Thus, a combination of the electric and Lorentz forces shifts the linear dispersion by $(u-qv)$ for the R-movers and by $(u+qv)$  for the L-movers. This results not only in different spacings between the $R$ and $L$ modes in Eq. (\ref{eq3}), but also in different localization properties of the corresponding wave functions. The wave functions (\ref{eq7}) are localized in the $z$ direction on a finite number of layers of the order of $t/|u \mp qv|$. So, with an increase of the electric field, the L (R) movers become more (less) localized. At the resonance condition $u=qv$, the R-movers experience zero net force, so their wave functions become delocalized, and the spectrum acquires a tight-binding dispersion
\begin{equation}  \label{eq8}
	\varepsilon^R=vp_x+ t\cos(k_z),   
\end{equation}
where the out-of-plane momentum $k_z$ is a good quantum number.

\begin{figure}
\centering	
\resizebox{0.9\columnwidth}{!}{\includegraphics{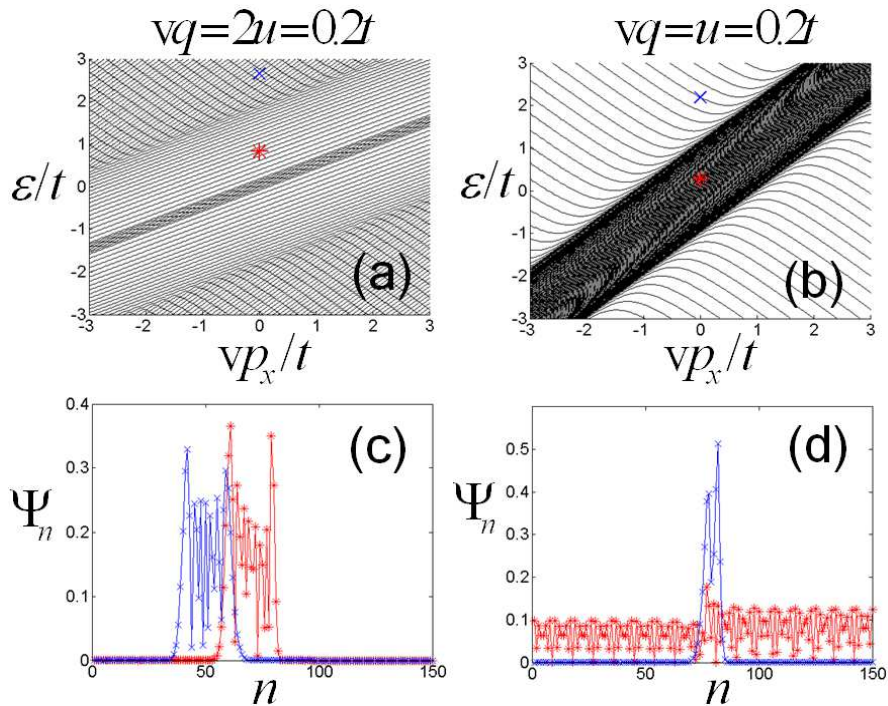}}
\caption{Numerically calculated energy spectra and wave functions for a finite system of 150 layers. The top panels show the energy spectra, whereas the bottom panels show the wave functions corresponding to the states marked by the (*) and (x) symbols in the top panels. The left and right columns correspond to the non-resonant $2u=qv$ and resonant $u=qv$ cases, respectively. }
\label{fig:6}     
\end{figure}

The numerically calculated energy spectra of the full Eq. (\ref{eq4}) are shown in Fig.~\ref{fig:6}(a) and (b), whereas the corresponding wave functions are shown in Fig.~\ref{fig:6}(c) and (d). Although details of the spectra differ from the simple description given in Eq. (\ref{eq3}), the general structure remains similar. For the case $u=0$ and $q\neq 0$, which was studied in Ref. \cite{ref7}, the spectrum consists of the discrete Landau levels within the energy window $|\varepsilon|<2t$ and a quasi-continuous spectrum outside this window $|\varepsilon|>2t$.  When a finite but non-resonant electric field $2u=qv=0.2t$ is applied, the discrete Landau levels acquire dispersion in $p_x$, as shown in Fig.~\ref{fig:6}(a). The quasi-continuous spectrum consists of a series of parallel lines corresponding to the L and R modes, well described by Eq. (\ref{eq3}). The wave functions are localized in the $z$ direction both for the discrete Landau levels and the quasi-continuous spectrum, as shown in Fig.~\ref{fig:6}(c). However, when the resonance between the fields is achieved at $u=qv=0.2t$, the spectrum changes drastically, as shown in Fig.~\ref{fig:6}(b). A continuous band of the R-movers corresponding to spectrum (\ref{eq8}) emerges, and the eigenstates become delocalized, as shown in Fig.~\ref{fig:6}(d). At the same time, the L-movers remain well-localized.

We believe that the mechanism of resonant delocalization of the wave functions in the crossed electric and magnetic fields, as discussed above, is responsible for the peaks in the differential conductance observed in the experiment. According to Eq. (\ref{eq1}), the position of the conductance peak is proportional to the magnetic field
\begin{equation}  \label{eq9} 
	E=V_0/l=vH,
\end{equation}
where $l$ is the effective length over which the applied voltage $V_0$ drops. Equation (\ref{eq9}) is consistent with the experimental Fig.~\ref{fig:3}, from which we obtain $l=1.2$ nm, about three times longer than the interlayer spacing in graphite. This result indicates that decoupling of the interlayer Bernal correlation probably occurs in two or three intrinsic tunnel junctions. In NbSe$_3$ mesas, it was demonstrated that the charge-density-wave decoupling can occur within two intrinsic tunnel junctions of the mesa \cite{ref9}. Although our theoretical model was developed for a large number of graphene layers, the resonant condition (\ref{eq1}) is expected to be valid even for a few layers. 

\section{Conclusion} \label{sec:3}

The interlayer tunneling spectra $dI/dV$ measured in graphite mesas subjected to a strong in-plane magnetic field $H$ demonstrate a peak at the voltage $V_0$, which is proportional $H$. The experimental result is consistent with the presented theory, where the wave functions of the 2D Dirac electrons in graphene layers delocalize in the out-of-plane direction when the resonant condition $E=vH$ between the fields is achieved.

\begin{acknowledgement}
The work was supported by RFBR grants No 11-02-01379-a, No 11-02-90515-Ukr\_f\_a, 11-02-12167-ofi-m, programs of RAS, by European Commission from the 7$^{\rm th}$ framework program ``Transnational access'', contract No 228043 ``Euromagnet II'' Integrated Activities, and partially performed in the frame of the CNRS-RAS Associated International Laboratory between Institut Neel and IRE ``Physical properties of coherent electronic states in condensed matter''.
\end{acknowledgement}

\end{document}